\documentclass[aps,prb, twocolumn, showpacs,superscriptaddress,dvipsnames]{revtex4-1}

\usepackage{amssymb,amsfonts,amsmath} 
\usepackage{graphicx,epsfig,psfrag}
\usepackage{color}
\usepackage{natbib}
\usepackage{url}
\usepackage[breaklinks=true]{hyperref}
\usepackage{mathtools}
\usepackage{subfigure}
\hypersetup{
        colorlinks=true,
        citecolor    = blue
}
\usepackage{bookmark}
\usepackage{epic}
\usepackage{longtable}

\usepackage{bbm}
\usepackage{ulem}
\usepackage{braket}
\usepackage{xcolor}
\usepackage{csquotes}
\usepackage{graphicx}
\usepackage{tikz}
\usepackage{extarrows}
\usetikzlibrary{trees}
\usetikzlibrary{decorations.pathmorphing}
\usetikzlibrary{decorations.markings}
\usetikzlibrary{positioning,arrows}
\usetikzlibrary{shapes.arrows}
\usepackage{leftidx}
\usepackage{framed}
\usepackage{tcolorbox}
\usepackage{bm}

\tikzset{->-/.style={decoration={
  markings,
  mark=at position #1 with {\arrow{>}}},postaction={decorate}}}

\newcommand{\tens}[1]{%
  \mathbin{\mathop{\otimes}\limits_{#1}}%
}

\begin{document} 

\title{Nonequilibrium thermoelectric transport through vibrating molecular quantum dots}

\author{A.\ Khedri}
\affiliation{Institut f{\"u}r Theorie der Statistischen Physik, RWTH Aachen University 
and JARA---Fundamentals of Future Information
Technology, 52056 Aachen, Germany}
\affiliation{Peter Gr\"unberg Institut and Institute for Advanced Simulation, Research Centre J\"ulich, 
52425 J\"ulich, Germany} 
\author{T.A.\ Costi}
\affiliation{Peter Gr\"unberg Institut and Institute for Advanced Simulation, Research Centre J\"ulich, 
52425 J\"ulich, Germany}
\author{V.\ Meden} 
\affiliation{Institut f{\"u}r Theorie der Statistischen Physik, RWTH Aachen University 
and JARA---Fundamentals of Future Information
Technology, 52056 Aachen, Germany}
\begin{abstract} 

We employ the functional renormalization group to study the effects of phonon-assisted tunneling on the nonequilibrium steady-state transport
through a single level molecular quantum dot coupled to electronic leads.
Within the framework of the spinless Anderson-Holstein model,
we focus on small to intermediate electron-phonon couplings, 
and we explore the evolution from the adiabatic to the antiadiabatic limit and also from the low-temperature non-perturbative regime to the high temperature perturbative one.
We identify the phononic signatures in the bias-voltage dependence of the electrical current and the differential conductance.
Considering a temperature gradient between 
the electronic leads, 
we further investigate the interplay between the transport of charge and heat.
Within the linear response regime, we compare the temperature dependence of various thermoelectric coefficients
to our earlier results obtained within the numerical renormalization group [Phys.~Rev.~B {\bf 96}, 195156 (2017)].
Beyond the linear response regime, in the context of thermoelectric generators, we discuss the influence of molecular vibrations on the output power and the efficiency.
We find that in the antiadiabatic limit the thermoelectric efficiency can be significantly enhanced.

\end{abstract}
\date{\today} 
\maketitle
\section{Introduction}
\label{sec:intro}

In molecular electronics,\cite{Galperin07} many-body correlations are expected to play an important role at low temperatures.\cite{Hewson97,Paaske05}
Transport measurements in such systems can reveal the direct consequences of the local Coulomb interaction or the local electron-phonon coupling.\cite{Qiu04}
Of particular interest is to study how these interactions modify the interplay between electrical and heat conduction.
Such studies can provide guidance for potential routes towards the use of such nano-structures for possible applications in on-chip cooling and waste-heat conversion.\cite{Mahan1998}
Theoretical guidance is needed to properly interpret experimental results.\cite{Linke17}
Improving the theoretical understanding of thermoelectric transport through vibrating molecular quantum dots
can thus help to address fundamental questions on the restrictions of the efficiency of molecular quantum dot-heat engines, and on the role of many-body effects in coupled irreversible processes (heat and electrical conduction).

In this work we consider the spinless Anderson-Holstein model (SAHM) which is a simple model to capture the effects of local vibrational degrees of freedom in molecular devices.\cite{Mitra04} 
The electron-phonon interaction results in the emergence of a low-energy scale $\Gamma_{\rm eff}$ which is smaller than the bare tunneling rate $\Gamma$ (see below for exact definitions).
In several studies it was investigated how this low-energy scale evolves with increasing the electron-phonon coupling, and the phonon frequency,
and a combined polaronic and x-ray edge like renormalization was found.\cite{Sherrington75,Hewson80,Eidelstein13,Khedri17}
Here our focus is on 
correlation effects in
the nonlinear thermoelectric transport properties of the SAHM. 
In the antiadiabatic limit, namely when the phonon frequency is much larger than the bare tunneling rate,
the physics is non-perturbative and we need sophisticated many-body methods.
For this purpose, we use the functional renormalization group (FRG),\cite{Metzner12} which goes well beyond perturbation theory even within its simplest approximation scheme (first-order truncation).

The FRG approach is a flexible theoretical method to tackle interacting many-body systems from two or one dimensional interacting fermionic models\cite{Metzner12}
to quantum impurity systems.\cite{Karrasch10t}
In particular, the effects of the local Coulomb interaction on the impurity site has been investigated within the single impurity Anderson model in equilibrium,\cite{Karrasch06} as well as in the steady-state nonequilibrium,\cite{Jakobs10}
and the influences of the nearest neighbor Coloumb interaction at the contact point between the impurity and the fermionic leads has been studied within the interacting resonant level model (IRLM) both in and out of equilibrium.\cite{Karrasch10}
The interplay between the local Coulomb interaction (Kondo physics) and the local electron-phonon coupling (polaron physics) has been also considered in the framework of the Anderson-Holstein model
within\cite{Kennes14} and beyond\cite{Laakso14} the linear response regime.
In contrast to previous applications of the FRG method, for the SAHM the bare interaction is frequency dependent (stemming from integrating out phonons, see below)
resulting in the frequency dependence of the self-energy even in the lowest order truncation scheme. 
In this work we extend our previous studies of the SAHM\cite{Khedri17,Khedri17b} by setting up the FRG approach on the Keldysh contour.\cite{Gezzi07,Jakobs07}
The latter allows us to compute the spectral function without the need for the analytic continuation,
and hence various linear thermoelectric transport coefficients, as well as to study the nonlinear bias-voltage and temperature gradient dependence of charge and energy currents.
In the linear response regime we compare the FRG results to those obtained within the numerical renormalization group (NRG) approach.\cite{Khedri17b}
As the nonequilibrium steady-state NRG is a method still in development,\cite{Anders08,Hoa14,Nghiem17} requiring the solution of the issues with thermalization at long times,\cite{Rosch12,Guettge13}
we do not employ NRG beyond the linear response regime. However see Ref.~\onlinecite{Nghiem18} for recent progress on overcoming these issues.

Two distinct signatures of phonon-assisted tunneling in nonequilibrium reported on in several experimental works,\cite{Park00,Zhitenev02,Qiu04,LeRoy04,Pasupathy04,Chae05,Sapmaz06,Leturcq09}
are the blockade of the charge current at small bias voltages (compared to the gate voltage $\tilde{\varepsilon}_0$),
and the vibrational excitations appearing as (approximately) equally-spaced peaks in the bias-voltage dependence of the differential conductance.
The first signature, the suppression of the electrical current for strong electron-phonon couplings,
and for small bias voltages ($\ll\text{max}\{\Gamma_{\rm eff},|\tilde{\varepsilon}_0|\}$, see below) has been theoretically verified for the SAHM.\cite{Huetzen12,Jovchev13,Koch11,koch12}
In particular, in Ref.~\onlinecite{Huetzen12}, 
the iterative summation of path integrals (ISPI) has been used, which is valid for sufficiently high temperatures or large bias voltages,
and in Ref.~\onlinecite{Jovchev13}, the scattering-states numerical renormalization group (SNRG) has been employed, with the exclusive focus on strongly asymmetric coupling to different reservoirs.
This suppresses the bias-voltage dependency of the spectral function and hence makes the calculations more feasible.  
The reduction of current has been denoted as the Franck-Condon blockade,
and was experimentally observed in suspended carbon nanotube quantum dots with longitudinal stretching mode (frequency $\approx 0.5~\text{meV}$)
in which sizable electron-phonon couplings (of the order of frequency) can be achieved,\cite{Leturcq09}
as well as carbon-based molecular transistors (e.g., $\text{C}_{60}$, and $\text{C}_{140}$).\cite{Park00,Pasupathy04}
This blockade is due to the formation of a massive local polaron and can be understood in terms of the equilibrium renormalized tunneling rate $\Gamma_{\textrm{eff}}$.
The second vibrational signature is the step-like I-V characteristic,\cite{Park00,Sapmaz06} resulting in multiple phononic peaks in the differential conductance.\cite{Qiu04,Chae05}
This has also been theoretically clarified for the SAHM in Ref.~\onlinecite{Koch11}, using an approach based on the variational Lang-Firsov transformation,
and in Ref.~\onlinecite{Schinabeck18}, employing a hierarchical quantum master equation approach.
Similar features have been found in the spinful version of the Anderson-Holstein model.\cite{Lueffe08,Leijnse08}
For this model the study of the current fluctuations in the strong Coulomb blockade regime has revealed avalanche-like transport of electrons
for strong electron-phonon couplings in the weak tunneling limit $\Gamma\ll T$, resulting in giant Fano factors.\cite{koch05}


Within the Keldysh FRG approach, besides reproducing the results from the aforementioned studies, 
by considering a finite temperature gradient, we further study the energy current as a function of arbitrary bias-voltage and temperature.
In our approach the tunneling processes between the dot and the leads are included to all orders.
Being bound to weak to intermediate electron-phonon coupling, we complement
the recent study Ref.~\onlinecite{Loos17}.
We characterize the nonlinear thermoelectric effects and we further identify situations for which these nonlinear effects can enhance the efficiency
in the context of thermoelectric generators, converting waste heat into electrical energy, by i.e., charging a nanoscale battery.
In an earlier study of the IRLM,
many-body effects due to the short-range Coulomb interaction were found to have nontrivial consequences resulting in the enhancement of the efficiency.\cite{Kennes13}
In this paper we analyze the influence of correlation effects induced by the molecular vibrational degrees of freedom.
The latter results in inelastic scattering processes leading to the dissipation of energy in the molecule, as has been discussed for the linear response regime, in Ref.~\onlinecite{Imry14} using perturbation theory.
We extend such an analysis to the nonlinear and nonperturbative regime.

We organize this paper as follows. First in Sec.~\ref{sec:Method}, we briefly introduce the SAHM, and the formal details of employing FRG on the Keldysh contour. Within the first-order truncation,
we obtain a set of coupled differential equations for various components of the molecular self-energy. Our results are presented in Sec.~\ref{sec:results}.
First, in the absence of a temperature gradient, we characterize the vibrational features in the bias-voltage dependency of the charge current and differential conductance.
We further investigate the performance of the molecular quantum-dot heat engines.
Finally, in Sec.~\ref{sec:conclusion}, we present a short summary and perspective.
In the Appendix we discuss the evolution of the nonequilibrium spectral and distribution function upon increasing the bias voltage.
Furthermore, we compare the FRG results for the molecular spectral function to the NRG for vanishing bias voltage. 
\section{Model and method}
\label{sec:Method}
The Anderson-Holstein model is defined by the Hamiltonian
\begin{align}
H = &\sum_{\alpha=\text{L,R}} \sum_k \left(\varepsilon_k-\mu_{\alpha}\right) c_{\alpha,k}^\dag c_{\alpha,k}\nonumber\\
&+\frac{1}{\sqrt{N_{\rm sites}}} \sum_{\alpha=\text{L,R}}t_{\alpha}\sum_k \left( d^\dag c_{\alpha,k} + \mbox{H.c.}  \right) \nonumber\\
&+\epsilon_0 d^\dag d + \omega_0 b^\dag b + \lambda d^\dag d (b^\dag +b).
\label{eq:Method_modelhamiltonian}
\end{align}
It features two leads of non-interacting electrons (ladder operators $c_{\alpha,k}^{(\dag)}$),
each of which is characterized by the chemical potential $\mu_{\alpha}$, and is represented by a one dimensional chain with $N_{\textrm{sites}}$ lattice sites, and dispersion $\epsilon_k$.
The leads  are coupled to a localized level (ladder operator $d^{(\dag)}$) with energy $\epsilon_0$ via tunneling processes with amplitude $t_{\textrm{L/R}}$.
The localized level is also coupled to a local vibrational mode with frequency $\omega_0$.
The coupling is such that the occupation of the molecule leads to a displacement of the oscillator.
The strength of this displacement can be tuned with $\lambda$, the electron-phonon coupling.
The particle-hole symmetric point of the Hamiltonian is $\epsilon_0=E_{\rm p}$, 
where $E_{\rm p}=\lambda^2/\omega_0$ is known as the polaronic shift.
Hence the quantity $\tilde{\varepsilon}_0=\epsilon_0-E_{\rm p}$ controls the charge on the molecule, and can be regarded as the gate voltage.

Not being interested in band effects, we consider the so called wide-band limit,
where the leads have a constant density of states $\rho_{\textrm{lead}}=1/(2D)$ in the interval $[-D,D]$, and $\rho_{\textrm{lead}}=0$ outside this interval.
Thereby the bare tunneling rate reads $\Gamma=\sum_{\alpha=\rm L,R}\Gamma_{\alpha}$,
with $\Gamma_{\alpha}=\pi\rho_{\textrm{lead}}t_{\alpha}^2$,
which determines the width of the resonance in single-particle tunneling 
in the absence of the electron-phonon coupling.
The electron-phonon coupling suppresses the rate of resonant tunneling and results in the appearance of phonon side peaks (see the Appendix). 
The former effect can be quantified by $\Gamma_{\rm eff}$ defined
via $\Gamma_{\rm eff}=1/(\pi \chi_c)$, where $\chi_c=-\frac{d n_{d}(\tilde{\varepsilon}_0)}{d\tilde{\varepsilon}_0}|_{\tilde{\varepsilon}_0=0}$
is the local $T=0$ charge susceptibility, and $n_d$ denotes the occupancy of the molecular level.




\subsection{Transport properties}
We assume that at time $t<t_0$ the localized level is decoupled from the vibrational degrees of freedom, and also from the electronic leads,
i.e., the system is uncorrelated with a density operator $\rho_{\rm eq}=\rho_{\rm L}\tens{ }\rho_{\rm R}\tens{ }\rho_{\rm d}\tens{ }\rho_{\rm b}$,
where $\rho_{\rm L(R)}$ represents the grand canonical density operator of the left(right) fermionic lead at temperature $T_{\rm L(R)}$, and chemical potential $\mu_{\rm L(R)}$,
$\rho_{\rm b}$ is the density operator of the single-mode bosonic bath at temperature $T_{\rm ph}$, and chemical potential $\mu_{\rm ph}=0$, and finally, $\rho_{\rm d}$ denotes the dot density operator.
At $t=t_0$ the electron-phonon interaction is turned on, and the molecule is coupled to the leads.
We are generally interested in the electrical and heat currents passing through the molecule for $t>t_0$
for a given temperature gradient $\Delta T=T_{\rm R}-T_{\rm L}$, and a bias voltage $V=(\mu_{\rm L}-\mu_{\rm R})/e$, with $e$ being the electric charge.
The charge, energy, and heat currents leaving the $\alpha$-th reservoir are defined as
$J^{\rm c}_{\alpha}=-e\langle\langle \partial_t [\hat{c}^{\dag}_{\alpha,k}(t) \hat{c}_{\alpha,k}(t)]\rangle\rangle$,
$J^{\rm E}_{\alpha}=\langle\langle \partial_t [\sum_{k}\epsilon_k \hat{c}^{\dag}_{\alpha,k}(t) \hat{c}_{\alpha,k}(t)]\rangle\rangle$,
and $J^{\rm Q}_{\alpha}=J^{\rm E}_{\alpha}+(\mu_{\alpha}/e)J^{\rm c}_{\alpha}$, respectively.
As we shall see these currents can be written in terms of the molecular propagator with the retarded (advanced) component defined as
\begin{align}
G^{\rm R}_{\rm mol}(t,t^\prime)&=\left[G^{\rm A}_{\rm mol}(t,t^\prime)\right]^*\nonumber\\
&=-i\Theta(t-t^\prime)\left\langle\left\langle \left\{\hat{d}(t),\hat{d^{\dag}}(t^\prime)\right\}\right\rangle\right\rangle,
\end{align}
and the Keldysh one as
\begin{align}
G^{\rm K}_{\rm mol}(t,t^\prime)=-i\left\langle\left\langle \left[\hat{d}(t),\hat{d}^{\dag}(t^\prime)\right] \right\rangle\right\rangle,
\end{align}
where \enquote{$\hat{~}$} refers to the Heisenberg picture, and $\langle\langle\cdots\rangle\rangle=\text{Tr}\{\cdots \rho_{\rm eq}\}$.

Integrating out the structureless leads, and being interested in the steady-state limit $t_0\to -\infty$ (assuming that the limit exists due to the presence of reservoirs), we can directly work in (single) frequency space
\begin{align}
G^{\rm R/A}_{\rm mol}(\omega)=[\omega -\epsilon_0\pm i\Gamma -\Sigma^{\rm R/A}(\omega)]^{-1},
\label{eq:method_GRA}
\end{align}
and
\begin{align}
G^{\rm K}_{\rm mol}(\omega)=\left\{\Sigma^{\rm K}(\omega)-2i\Gamma\left[1-2f_{\rm eff}(\omega)\right]\right\}G^{\rm R}_{\rm mol}(\omega)G^{\rm A}_{\rm mol}(\omega),
\label{eq:method_GK}
\end{align}
with $\Sigma^{\rm R/A/K}(\omega)$ being the components of the molecular self-energy resulting from the presence of electron-phonon coupling,
$f_{\rm eff}(\omega)=\sum_{\alpha=\rm L,R}(\Gamma_{\alpha}/\Gamma)f_{\alpha}(\omega)$,
with the Fermi function $f_{\alpha}(\omega)=\left[\exp\{\beta_{\alpha}(\omega-\mu_{\alpha})\}+1\right]^{-1}$,
and the inverse temperature $\beta_{\alpha}=1/(k_{\rm B} T_{\alpha})$,
$k_{\rm B}$ being the Boltzmann constant.
The symmetric charge current $J^{\textrm{c}}=(\Gamma_{\rm R}/\Gamma)J^{\rm c}_{\textrm{L}}-(\Gamma_{\rm L}/\Gamma)J^{\rm c}_{\textrm{R}}$ reduces to
\begin{align}
J^{\textrm{c}}
&=\frac{e}{h}\frac{4\pi\Gamma_{\textrm{L}}\Gamma_{\textrm{R}}}{\Gamma}\int_{-\infty}^{\infty}d\omega[f_{\textrm{L}}(\omega)-f_{\textrm{R}}(\omega)]A(\omega),
\label{eq:Method_charge_current_sym}
\end{align}
where $A(\omega)=(-1/\pi)\text{Im}\left\{G_{\textrm{mol}}^{\textrm{R}}(\omega)\right\}$ represents the spectral function, and $h$ denotes the Planck's constant.
The energy current entering each reservoir reads
\begin{align}
J^{\textrm{E}}_{\textrm{L/R}}=\frac{-4\pi}{h}\Gamma_{\textrm{L/R}}\int_{-\infty}^{\infty}d\omega \omega [f_{\rm L/R}(\omega)-f_{\textrm{NE}}(\omega)]A(\omega),
\label{eq:NTT_heatcurrent}
\end{align}
with the nonequilibrium distribution function defined as
\begin{align}
f_{\rm NE}(\omega)= \frac{1}{2}\left\{1-\frac{G^{\rm K}_{\rm mol}(\omega)}{G^{\rm R}_{\rm mol}(\omega)-G^{\rm A}_{\rm mol}(\omega)}\right\}.
\label{eq:non_dis}
\end{align}
For the symmetric tunneling $\Gamma_{\textrm{R}}=\Gamma_{\textrm{L}}$, the difference of the heat currents entering the left and right leads reduces to
\begin{align}
 J^{\textrm{Q}}_{\textrm{R}}- J^{\textrm{Q}}_{\textrm{L}}=\frac{-2\pi\Gamma}{h}\int_{-\infty}^{\infty}d\omega \omega[f_{\textrm{R}}(\omega)-f_{\textrm{L}}(\omega)]A(\omega).
\label{eq:heat_current_difference}
\end{align}
In the absence of a temperature gradient $\Delta T=0$, and at particle-hole symmetry $\tilde{\varepsilon}_0=0$, the integrand is odd and hence the difference vanishes.
In other words, the heat current enters each reservoir symmetrically $ J^{\textrm{Q}}_{\textrm{L}}= J^{\textrm{Q}}_{\textrm{R}}$.


In general, as we have an energy-conserving system, it holds that
\begin{align}
\langle\langle\partial_t \hat{H}(t)\rangle\rangle=J^{\rm E}_{\rm R}+J^{\rm E}_{\rm L}+ \dot{E}_{\textrm{mol}}=0,
\label{eq:energyconservation}
\end{align}
with the molecular energy dissipation rate
\begin{equation}
\dot{E}_{\rm mol}\equiv \langle\langle\partial_t [\hat{H}_{\rm coup}(t)+\hat{H}_{\rm mol}(t)]\rangle\rangle,
\label{eq:Emoldef}
\end{equation}
where $H_{\rm coup}$ is the coupling term [the second line of Eq.~(\ref{eq:Method_modelhamiltonian})],
and $H_{\rm mol}$ is the molecular contribution to the Hamiltonian [the last line of Eq.~(\ref{eq:Method_modelhamiltonian})].
Equivalently, in terms of the charge and heat currents, the energy dissipation rate reads
\begin{align}
\dot{E}_{\textrm{mol}}&=-J^{\rm Q}_{\rm L}-J^{\rm Q}_{\rm R}+\frac{\mu_{\textrm{L}}-\mu_{\textrm{R}}}{e} J^{\textrm{c}}\\
&=\frac{4\pi\Gamma}{h}\int d\omega \omega [f_{\rm eff}(\omega)-f_{\textrm{NE}}(\omega)]A(\omega).
\label{eq:energy_dissipation_rate}
\end{align}
In the absence of electron-phonon coupling, $f_{\rm NE}(\omega)$ reduces to $f_{\rm eff}(\omega)$ [see Eq.~\ref{eq:non_dis}],
and hence the molecular dissipation rate vanishes, i.e., the energy is only being exchanged between the electronic leads.
In other words, the molecular dissipation rate vanishes
in the steady state limit in the absence of a coupling to the phonon mode.
However, for $\lambda\neq 0$, inelastic scattering processes (frequency-dependent self-energy) induced by phonon-assisted tunneling
can modify the noneqilibrium distribution function, potentially implying the dissipation of energy, i.e.,
energy can be exchanged not only between the fermionic leads but also with the phonon bath.
Note that the molecular dissipation rate $\dot{E}_{\rm mol}$ as defined in Eq.~(\ref{eq:energyconservation}), includes any form of energy
not being dissipated as heat in the electronic leads. It contains two contributions:
the expectation value of the molecular Hamiltonian as well as that of the molecule-lead coupling part [see Eq.~(\ref{eq:Emoldef})].
\subsection{The FRG approach to the SAHM}
\label{subsec:Method_flow}
Employing the functional integral formulation, the partition function of the SAHM is represented by an integral over both the fermionic and the bosonic fields.
However, 
we can integrate out the vibrational degrees of freedom (bosons), and obtain an effective (purely fermionic) action with a local in space while non-local in time (retarded)
two-particle interaction which in the tridiagonal representation \cite{Rammer07b} takes the form
\begin{align}
\tilde{u}&=\lambda^2
\begin{pmatrix}
D^{\rm R} (t,t^\prime) & D^{\rm K}(t,t^\prime)\\
0 & D^{\rm A}(t,t^\prime)\\
\end{pmatrix},
\end{align}
with the phonon propagator defined via
\begin{align}
&D^{\rm R} (t,t^\prime)=\left[D^{\rm A} (t,t^\prime)\right]^*=-i\Theta(t-t^\prime)\langle\langle[\hat{A}(t),\hat{A^{\dag}}(t^\prime)]\rangle\rangle ,\\
&D^{\rm K} (t,t^\prime)=-i\langle\langle\{\hat{A}(t),\hat{A^{\dag}}(t^\prime)\}\rangle\rangle ,
\end{align}
with $A\equiv b+b^{\dag}$. Therefore in frequency space we obtain
\begin{align}
&D^{\rm R} (\omega)=\left[D^{\rm A} (\omega)\right]^*=\frac{2\omega_0}{(\omega+i\eta)^2-\omega_0^2}, \\
&D^{\rm K}(\omega)=-2\pi i \left[1+2b(\omega_0)\right]\sum_{s=\pm} \delta(\omega-s\omega_0),
\end{align}
where $\eta\to 0^{+}$ has been introduced to guarantee convergence, and $b(\omega)=[\exp\{\beta_{\rm ph}\omega\}-1]^{-1}$ denotes the Bose distribution function at temperature $T_{\rm ph}=1/(k_{\rm B}\beta_{\rm ph})$.
Therefore the molecular vibrations result in a frequency-dependent (bare) interaction in the fermionic action. 

In the FRG approach,\cite{Metzner12} a flow parameter $\Lambda \in [0,\infty]$ is introduced (in the free molecular propagator) and 
high-frequency degrees of freedom (compared to $\Lambda$) are being integrated out.
From this procedure, we can obtain a hierarchy of differential equations (flow equations) for the one-particle irreducible vertex functions, e.g., the self-energy, and the effective two-particle interaction.
Truncation schemes are required to keep the calculations manageable.
Here, we focus on the first-order truncation scheme
(controlled for weak to intermediate electron-phonon couplings), i.e., only the self-energy flows as we change the flow parameter from $\infty$ to $0$.
This scheme has been successfully employed for the IRLM in and out of equilibrium,\cite{Karrasch10} or even for explicit time dependencies,\cite{Kennes12}
and also for the SAHM in equilibrium (finding good agreement with the nonperturbative NRG approach).\cite{Khedri17}

To obtain the flow equation for the various components of the self-energy, we need to specify a scale-dependent free molecular propagator.
We use the reservoir cutoff, as proposed in Ref.~\onlinecite{Jakobs09},
having the advantage of preserving some symmetries (like causality)
even at the lowest order truncation.\footnote{This scheme has been chosen to avoid the undesired artifacts of the sharp cutoff scheme, see Refs.~\onlinecite{Gezzi07,Jakobs07}.}
In this scheme, one assumes that at each $\Lambda$, the molecular level is coupled to an auxiliary reservoir with tunneling rate $\Lambda$, and distribution function $f_{\rm eff}(\omega)$.
Hence, the scale-dependent propagator $G^{\Lambda,\textrm{R/A/K}}_{{\textrm{mol}}}(\omega)$ can be determined analogous to Eqs.~(\ref{eq:method_GRA}) and (\ref{eq:method_GK}),
simply by replacing $\Gamma$ by $\Gamma+\Lambda$, and $\Sigma^{s}(\omega)$ by $\Sigma^{\Lambda,\textrm{s}}(\omega)-\epsilon_0$, with $s=\rm R,A,K$.
In this way, all energy scales can be addressed, and the infrared divergences that often show up in perturbation theory can be regularized.
Following the standard procedure, we obtain the flow equations as
\begin{align}
\partial_{\Lambda}\Sigma^{\Lambda,\textrm{R/A}}(\nu)=&-i\frac{\lambda^2}{4\pi}\bigg\{ \frac{2}{\omega_0}\int_{-\infty}^{\infty}d\omega S^{\Lambda,\textrm{K}}(\omega)\nonumber\\
&+\sum_{s\neq s^\prime=\rm R/A,K}
\int_{-\infty}^{\infty}d\omega S^{\Lambda,s}(\omega)D^{s^\prime}(\nu-\omega)\bigg\},
\label{eq:flowRA}
\end{align}
\begin{align}
\partial_{\Lambda}\Sigma^{\Lambda,\textrm{K}}(\nu)=-i\frac{\lambda^2}{4\pi}\bigg\{&\frac{2}{\omega_0}\sum_{s=\rm R,A}\int_{-\infty}^{\infty}d\omega S^{\Lambda,s}(\omega)\nonumber\\
&+\sum_{s=\rm R,A,K}\int_{-\infty}^{\infty}d\omega S^{\Lambda,s}(\omega)D^{s}(\nu-\omega)\bigg\},
\label{eq:flowK}
\end{align}
with single-scale propagator
\begin{align}
S^{\Lambda,\text{R}}(\omega)=&i[G^{\Lambda,\textrm{R}}_{\textrm{mol}}(\omega)]^2=\big[S^{\Lambda,\textrm{A}}(\omega)\big]^*,
\label{eq:SLamRA}\\
S^{\Lambda,\textrm{K}}(\omega)=&iG^{\Lambda,\textrm{R}}_{\textrm{mol}}(\omega)G^{\Lambda,\textrm{K}}_{\textrm{mol}}(\omega)
-iG^{\Lambda,\textrm{K}}_{\textrm{mol}}(\omega)G^{\Lambda,\textrm{A}}_{\textrm{mol}}(\omega)\nonumber\\
&+G^{\Lambda,\textrm{R}}_{\textrm{mol}}(\omega)\bigg[2i\big[1-2f_{\rm eff}(\omega)\big]\bigg]G^{\Lambda,\textrm{A}}_{\textrm{mol}}(\omega). 
\label{eq:SLamK}
\end{align} 
One can show that particle-hole symmetry is preserved for any $\Lambda$. 
Furthermore, in thermal equilibrium,
the fluctuation dissipation theorem (FDT) is preserved during the flow.
We solve the coupled differential equations Eqs.~(\ref{eq:flowRA}) and (\ref{eq:flowK}), with initial conditions
\begin{align}
\Sigma^{\Lambda\to\infty,\rm R/A}(\nu)-\epsilon_0 =E_{\rm p}, \quad 
\Sigma^{\Lambda\to\infty,\rm K}(\nu)=0,
\label{eq:Method_initial}
\end{align}
numerically using standard adaptive routines.
We always checked the convergence of the results with respect to the frequency grid (technical necessity to solve the differential equations),
and the symmetries such as the particle-hole symmetry at $\epsilon_0=E_{\rm p}$, and the FDT at $eV=0$, and $\Delta T=0$ has been numerically verified up to machine precision.
\section{Results}
\label{sec:results}
Having access to the frequency structure of various components of the molecular Green's functions Eqs.~(\ref{eq:method_GRA}), and (\ref{eq:method_GK}), in Sec.~\ref{subsec:results_IV} we study the bias-voltage dependence of the 
charge current Eq.~(\ref{eq:Method_charge_current_sym}), and the differential conductance at different gate voltages.
In Sec.~\ref{subsec:IT}, we discuss the evolution of electrical current and energy dissipation rate as one enters the high-temperature regime from the low-temperature one.
Finally, in Sec.~\ref{subsec:thermo} we study the thermoelectric transport through a vibrating molecule trapped between leads held at different temperatures and chemical potentials.
In the linear regime we compare the transport coefficients with NRG, while in the nonlinear regime we discuss how the efficiency of a thermoelectric generator can be enhanced by vibrational degrees of freedom.
\begin{figure}[!htbp]
 \centering
\includegraphics[width=1.0\linewidth]{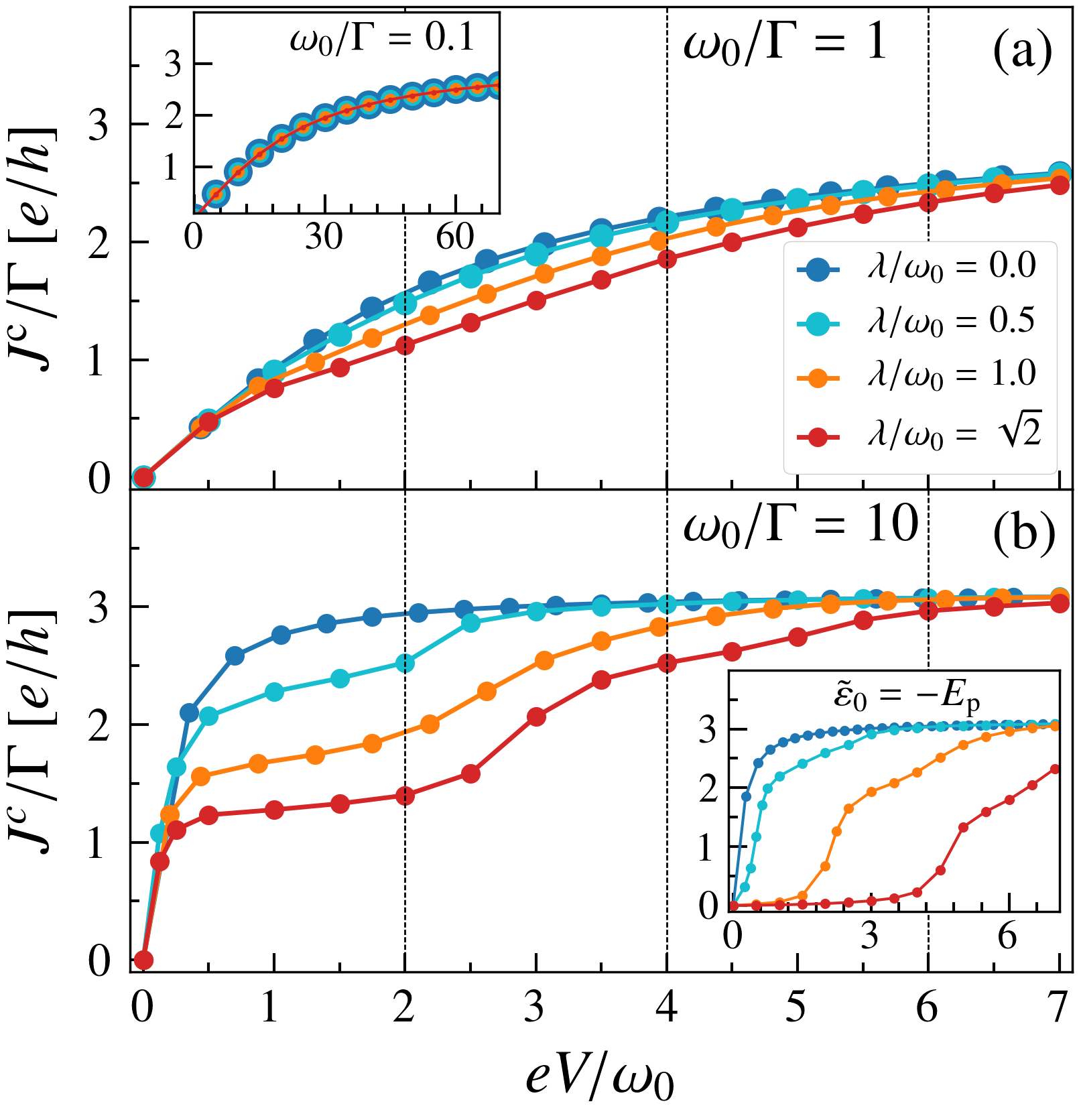}
 \caption{
Charge current $J^{\rm c}$ vs bias-voltage $V$ at particle-hole symmetry $\tilde{\varepsilon}_0=0$,
for the listed electron-phonon couplings $\lambda/\omega_0$,
(a), in the crossover regime of phonon frequency $\omega_0/\Gamma=1$, and,
(b), in the antiadiabatic limit $\omega_0/\Gamma=10$.
 Inset to (a): $J^{\rm c}$ vs $V$ in the adiabatic limit $\omega_0/\Gamma=0.1$.
 Inset to (b): $J^{\rm c}$ vs $V$ at finite gate voltage $\tilde{\varepsilon}_0=-E_{\rm p}$
 in the antiadiabatic limit ($\omega_0/\Gamma=10$).
 For $\lambda/\omega_0=0,0.5,1,\sqrt{2}$,
 we find $\Gamma_{\rm eff}/\Gamma=1,0.92,0.70,0.46$ for the crossover regime in (a),
 while for the antiadiabatic regime in (b), we find $\Gamma_{\rm eff}/\Gamma=1,0.83,0.45,0.19$.
 The temperature is $k_{\rm B}T/\Gamma=0.1$ in all plots.}
 \label{fig:IV}
\end{figure}
\begin{figure*}[t]
\centering
\includegraphics[width=1.0\linewidth]{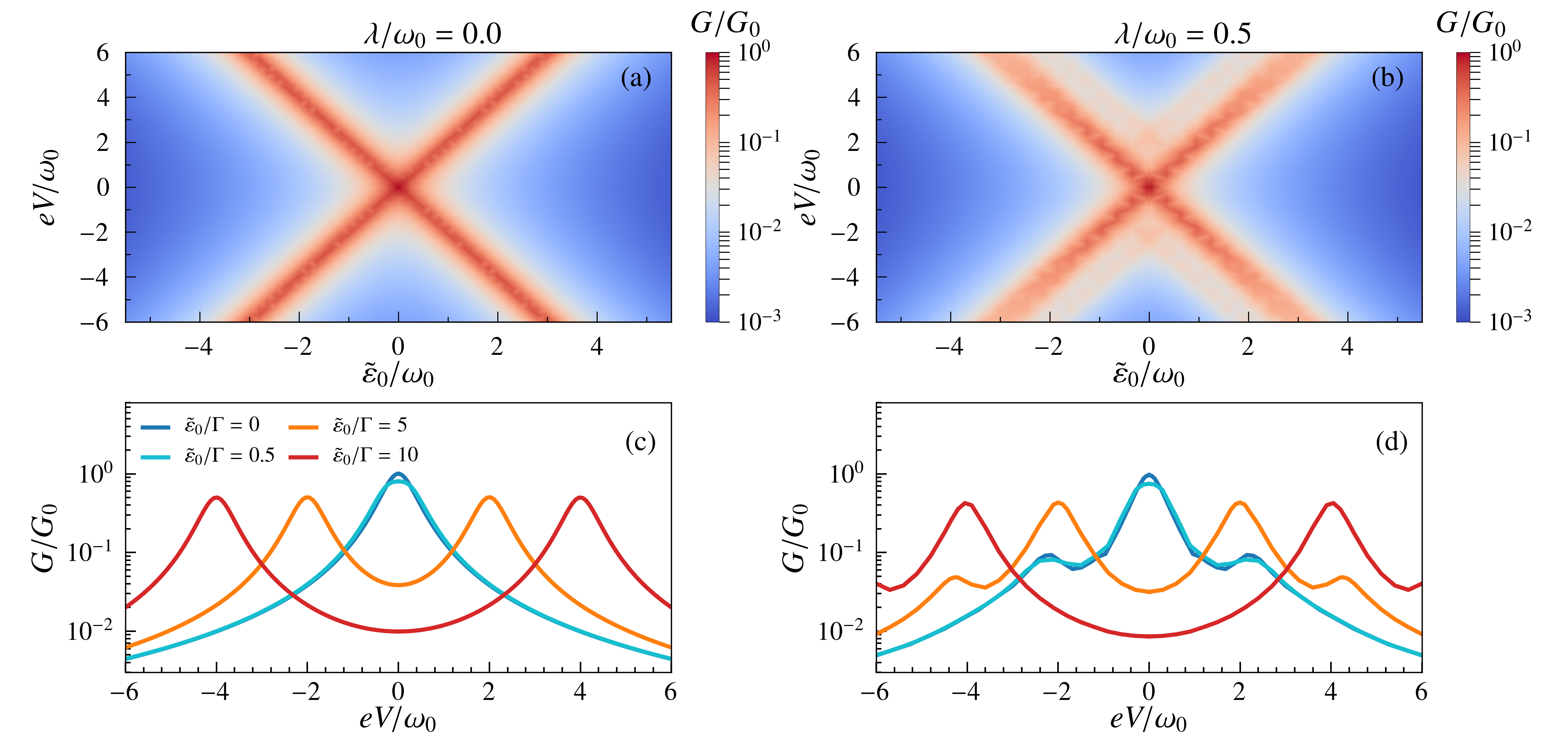}
\caption{(a) and (b) The differential conductance $G/G_0$ as a function of bias-voltage and gate voltage (stability diagram) in the antiadibnatic limit $\omega_0/\Gamma=5$, at $k_{\rm B}T/\Gamma=0.1$,
for (a) $\lambda/\omega_0=0$, and (b) $\lambda/\omega_0=0.5$. (c) and (d) The bias-voltage dependence of $G$ for various gate voltages for $\lambda/\omega_0=0,0.5$, respectively.}
\label{fig:Gd_stabil}
\end{figure*}
\subsection{The bias voltage dependence of the charge current and the differential conductance}
\label{subsec:results_IV}
We investigate the current-bias-voltage characteristic of the SAHM,
in the absence of a temperature gradient $T_{\textrm{L}}=T_{\textrm{R}}=T_{\textrm{ph}}=T$.
At the particle-hole symmetry, we consider the evolution with increasing the electron-phonon coupling from the adiabatic limit to the antiadiabatic one.
As we shall see the phonon signatures are more prominent in the antiadiabatic limit in which the charge fluctuations can effectively (de)excite the vibrational degrees of freedom
(as the dwell time of electrons on the molecule is larger than time-scale of the molecular vibrations $1/\Gamma\gg 1/\omega_0$).
Going beyond particle-hole symmetry, we further comment on the bias-voltage dependence of the current and the differential conductance in the antiadiabatic limit.

We first focus on the particle-hole symmetric case, $\tilde{\varepsilon}_0=0$.
In the adiabatic limit ($\omega_0/\Gamma=0.1$) the modification of the bias-voltage dependence of the charge current for different electron-phonon couplings is minor, see the inset of Fig.~\ref{fig:IV} (a).
In the crossover regime ($\omega_0\approx\Gamma_{\rm eff}$), see Fig.~\ref{fig:IV} (a), increasing the electron-phonon coupling the current is suppressed,
and at $eV\approx\omega_0$ an inflection point starts to form for stronger couplings.
In the antiadiabatic limit as illustrated in Fig.~\ref{fig:IV} (b), this suppression is more pronounced, and we see the development of multiple phonon steps as we approach the strong coupling regime. 
The sharp initial step is just a manifestation of the suppression of the tunneling
($\Gamma_{\textrm{eff}}$ as listed in the caption of Fig.~\ref{fig:IV}, for more see Refs.~\onlinecite{Eidelstein13}, and \onlinecite{Khedri17}), and
the additional steps reflect the possibility of charge transport through the molecule via inelastic processes, i.e., the absorption (emission) of one or multiple phonons.  
One might compare these results to those presented in Ref.~\onlinecite{Koch11} (qualitatively similar).
We may note that the steps do not occur exactly at integer multiples of $\omega_0$ but at $eV/2\gtrapprox \omega_0, 2\omega_0,\cdots$.
This shift can be interpreted as the renormalization of the phonon frequency as has been discussed in Ref.~\onlinecite{Eidelstein13}.
We conclude that in the antiadiabatic limit, the electronic degrees of freedom elevates the frequency of the molecular vibrations.

Finally, we show the effects of particle-hole asymmetry on the I-V characteristic for the antiadiabatic case ($\omega_0/\Gamma=10$) in the inset of Fig.~\ref{fig:IV} (b).
The main effect is that the current is blocked at low temperatures $eV\geq \text{max}\{\Gamma_{\rm eff},\tilde{\varepsilon}_0\}$ for sizable electron-phonon couplings.
Applying larger bias voltages can eventually lift up the blockade, and phonon steps will appear.


These phononic features have been reported in the experiments performed on suspended carbon nanotubes,\cite{Sapmaz06} appearing as an external structure on top of the Coulomb diamonds (spin-full version). However,
in the mentioned electronic transport spectroscopy measurements, the vibrational steps are sometimes accompanied with negative differential conductance,
which does not show up in our calculations for the SAHM in the parameter regime we have considered.

Figure~\ref{fig:Gd_stabil} illustrates the bias-voltage and the gate-voltage dependence of the differential conductance defined as $G=\frac{\partial J^c}{\partial V}|_{\Delta T=0}$.
For the resonant level model ($\lambda=0$), $G$ reads
\begin{align}
G=\frac{G_0}{2}\sum_{s=\pm}\frac{\Gamma^2}{[(seV/2)-\tilde{\varepsilon}_0]^2+\Gamma^2},
\label{eq:results_Gd_l0}
\end{align}
with $G_0=e^2/h$. Therefore at a given gate voltage $|\tilde{\varepsilon}_0|\geq\Gamma/\sqrt{3}$, the differential conductance exhibits two peaks, see Fig.~\ref{fig:Gd_stabil} (a) and \ref{fig:Gd_stabil} (c).
In the presence of molecular vibrations $\lambda\neq 0$, at low temperatures $k_{\rm B}T\ll\Gamma_{\rm eff}$, and low bias-voltages $eV\ll\omega_0$,
we can approximate the differential conductance analogous to Eq.~(\ref{eq:results_Gd_l0})
with replacing $\Gamma$ by $\Gamma_{\rm eff}$ [$=0.85\Gamma$ for the chosen parameters in Fig.~\ref{fig:Gd_stabil} (b) and \ref{fig:Gd_stabil} (d)].
This way we can understand the suppression of the differential conductance at finite gate voltages, see Fig.~\ref{fig:Gd_stabil}~(b).
Close to the particle-hole symmetric point,
we get a shoulder at $eV\approx\omega_0$ [see the curves corresponding to $\tilde{\varepsilon}_0/\Gamma=0,0.5$ in Fig.~\ref{fig:Gd_stabil}~(d)].
For larger bias-voltages $eV>\omega_0$ we obtain multiple phonon side peaks as shown in Fig.~\ref{fig:Gd_stabil}~(b) and
\ref{fig:Gd_stabil} (d). 
\subsection{Evolution with varying the temperature}
\label{subsec:IT}
\begin{figure}[!htbp]
 \centering
\includegraphics[width=1.0\linewidth]{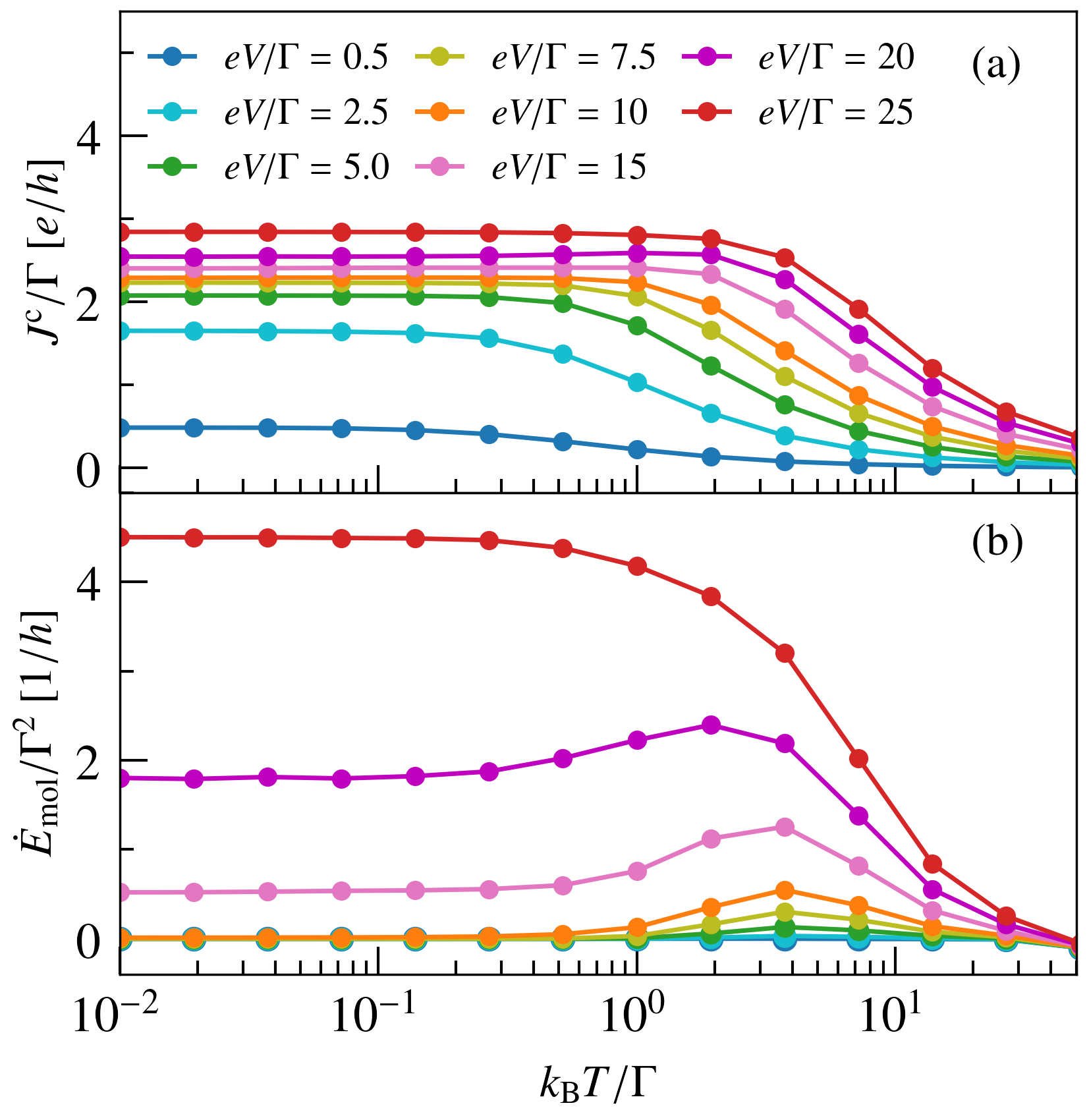}
 \caption{(a) Electrical current, and (b) molecular dissipation rate as a function of temperature for varying bias voltage and coupling $\lambda/\omega_0=0.5$,
 in the antiadiabatic limit $\omega_0/\Gamma=10$, and at particle-hole symmetry $\tilde{\varepsilon}_0=0$.}
 \label{fig:Tscan}
\end{figure}

In this section we elucidate on the modification of the charge and energy current while increasing the temperature ($T_{\rm L}=T_{\rm R}=T_{\rm ph}=T$). 
We focus on the particle-hole symmetric point for which the energy current enters each lead symmetrically [see discussion in connection to Eq.~(\ref{eq:heat_current_difference})].
As illustrated in Fig.~\ref{fig:Tscan} (a) and \ref{fig:Tscan} (b), for a given bias voltage, at low temperatures $k_{\rm B}T\ll\Gamma_{\textrm{eff}}$, both the electrical current and the molecular energy dissipation rate (which vanishes for $\lambda=0$) are independent of temperature.
This is the so called coherent transport regime as discussed in Ref.~\onlinecite{Galperin07}.
However, at higher temperatures $k_{\rm B}T>\max\{\Gamma_{\textrm{eff}},\tilde{\varepsilon}_0\}$, the electrical current decreases with increasing temperature, see Fig.~\ref{fig:Tscan} (a).
At such high bias voltages, there is a competition between resonant tunneling and phonon-assisted tunneling (satellite peaks are inside the bias window, see the Appendix).
If we increase the bias voltage further $eV>\omega_0$, the molecular energy dissipation rate exhibits a maximum at temperatures related to the phonon frequency, as shown in Fig.~\ref{fig:Tscan} (b). 
The latter indicates that at elevated temperatures ($\approx\omega_0/2$) the molecular vibrational degrees of freedom will be excited,
and hence the charge fluctuations are suppressed (reduction of current), while the energy dissipation rate is enhanced.
Eventually however $\dot{E}_{\rm mol}$ monotonically decreases as a function of temperature for $T\gg\Gamma_{\rm eff}$.
\subsection{Thermoelectric transport}
\label{subsec:thermo}
Finally, we investigate the interplay between transport of charge and heat in the presence of a temperature gradient as well as a bias voltage.
First within the linear response limit, we compare the FRG results with the NRG data presented in Ref.~\onlinecite{Khedri17b}.
This confirms that lowest-order FRG provides reliable results for small to intermediate $\lambda/\omega_0$.
Next, going beyond the linear response regime, we explore the parameter space to find regimes in which molecular vibrations can result in the enhancement of the efficiency
of thermoelectric generators operating in the nonlinear regime.
\subsubsection{Linear thermoelectric transport}
\label{subsec:results_Thermo_linear}
\begin{figure}[!htbp]
 \centering
\includegraphics[width=1.0\linewidth]{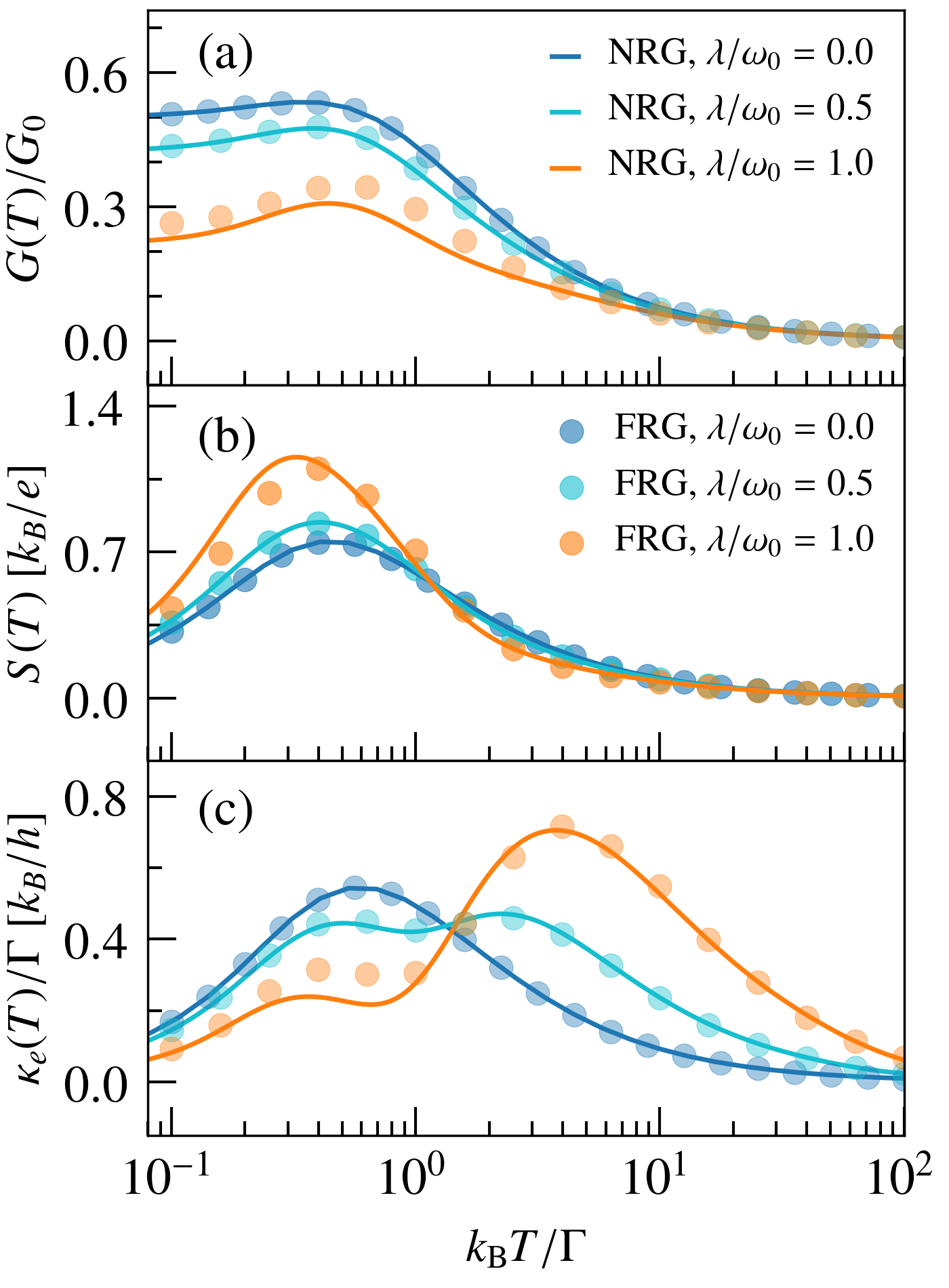}
 \caption{(a) The normalized electrical conductance $G/G_0$, (b) the Seebeck coefficient $S$ (in units of $k_{\textrm{B}}/e$), and (c) the normalized electronic contribution to the thermal conductance
 $\kappa_e/\Gamma$ (in units of $k_{\textrm{B}}^2/e^2$)
 vs reduced temperature for various electron-phonon coupling strengths
 in the antiadiabatic limit ($\omega_0/\Gamma=5.0$) and for a fixed gate voltage $\tilde{\varepsilon}_0/\Gamma=-1.0$.
 The solid lines represent the NRG data, and the circles are obtained within FRG by calculating the charge and heat current for an infinitesimal temperature gradient and bias voltage
 (calculating the transport coefficients by taking the derivatives numerically).}
 \label{fig:NTT_GSK_lscan}
\end{figure}

In the linear response regime, all the transport coefficients, i.e., the electrical conductance $G(T)=\frac{\partial J^{\rm c}}{\partial V}|_{\Delta T=0}$,
the Seebeck coefficient (thermopower) $S(T)=\frac{\partial V}{\partial \Delta T}|_{J^{\rm c}=0}$,
and the electronic contribution to the thermal conductance $\kappa_{\rm e}(T)=\frac{\partial J^{\rm Q}_{\rm R}}{\partial \Delta T}|_{J^{\rm c}=0}$,
can be expressed in terms of the moments of the molecular spectral function which can be accurately computed within the NRG approach.
Figure~\ref{fig:NTT_GSK_lscan} shows the comparison of taking the linear response limit of the FRG results (via calculating currents for $eV,k_{\rm B}\Delta{T}\ll\Gamma_{\rm eff},k_{\rm B}T$) to the NRG ones.
As shown, they match remarkably well at all temperatures and
only when approaching the strong coupling regime ($\lambda/\omega_0\gtrapprox1.0$), we see a small deviation in the temperature range $0.1\lesssim k_{\rm B}T/\Gamma\lesssim 1.0$.
We may note that the deviation of the FRG and NRG results for the thermal conductance is more pronounced for the low-temperature peak (resonant tunneling).
At such low temperatures one requires higher-order truncation schemes to capture the non-perturbative physics properly for strong electron-phonon couplings.
As discussed in our previous study,\cite{Khedri17b} the enhancement of the Seebeck coefficient together with the suppression of thermal conductance at low-temperatures
result in a sizable figure of merit $ZT_0=[TG(T)S^2(T)]/\kappa_{\rm e}(T)$, useful in harvesting waste heat.\cite{Mahan95}
\subsubsection{Nonlinear thermoelectric generator}
\label{subsec:results_Thermo_beyond}
\begin{figure}[!htbp]
\begin{tikzpicture}[scale=0.55] 

\fill[Tan!30!white] (0,0) ellipse (1.2cm and 1.2cm);

\filldraw[draw=NavyBlue,fill=NavyBlue, fill opacity=0.3,line width=.3mm] (-5.2,-1) rectangle (-2.8,3);

\draw [RedViolet,thick,domain=-45:45] plot ({1.5*cos(\x)}, {1.5*sin(\x)});
\draw [RedViolet,thick,domain=-30:30] plot ({2*cos(\x)}, {2*sin(\x)});

\draw [RedViolet,thick,domain=135:225] plot ({1.5*cos(\x)}, {1.5*sin(\x)});
\draw [RedViolet,thick,domain=150:210] plot ({2*cos(\x)}, {2*sin(\x)});


\draw[thick](-0.52,0.2)--(0.48,0.2);

\draw (-4.0,1.5) node[black]{\footnotesize Source};
\draw (-4.0,0.7) node[black]{\tiny $T_{\rm L}$};

\draw (4.0,0.5) node[black]{\footnotesize Drain};
\draw (4.0,-0.2) node[black]{\tiny $T_{\rm R}>T_{\rm L}$};

\filldraw[draw=BrickRed,fill=Red,fill opacity=0.3,line width=.3mm] (2.8,-1) rectangle (5.2,1.0);
\draw[->-=0.5,line width=.3mm,black,dashed] (-0.03,0.2)to [bend right=65](-4,3.0);
\draw[->-=0.5,line width=.3mm,black,dashed] (4,1.0)to [bend right=65](0.03,0.2);

\node at (2.1,0.0) [
    left color=white!70!Red,
    single arrow,
    shading angle=90+60,
    minimum height=1.2cm,
    minimum size=1.0cm, 
    rotate=-180
] { };

\node at (-1.75,0) [
    left color=white!70!Red,
    single arrow,
    shading angle=90+60,
    minimum height=1.2cm,
    minimum size=1.0cm, 
    rotate=-180
] { };

\node at (0.0,4.0) [
    left color=white!70!MidnightBlue,
    single arrow,
    minimum height=3.0cm,
    shading angle=90+60,
    rotate=-180
] { };
\draw (0.0,4.0) node[black]{\tiny $P>0$};
\draw (2.0,0) node[black]{\tiny $J^{\rm Q}_{\rm{R}}<0$};
\draw (-1.8,0) node[black]{\tiny $J^{\rm Q}_{\rm{L}}>0$};


\draw (0.0,-0.5)[black]node{\tiny $T_{\textrm{ph}}=T_{\rm L}$};

\fill[black] (0,0.2) ellipse (0.1cm and 0.1cm);
\fill[black] (-4,3) ellipse (0.1cm and 0.1cm);
\fill[black] (4,1) ellipse (0.1cm and 0.1cm);

\end{tikzpicture}
\caption{The sketch of a molecular quantum dot with vibrational degrees of freedom trapped between two electronic leads
held at different temperatures and chemical potentials.
The dashed lines indicate the transport of charge against the bias-voltage by means of the temperature gradient, i.e., charging of a battery.}
\label{fig:sketch}
\end{figure}
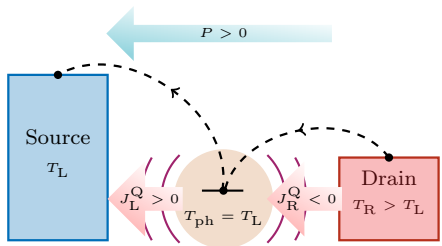
In this section we study the role of molecular vibrations in the performance of thermoelectric generators in nonequilibrium scenarios. 
We assume that the temperature and chemical potential of each reservoir are held fixed with the aid of some external energy source. 
Assuming $eV>0$, and $T_{\textrm{L}}=T_{\textrm{ph}}=T_{\textrm{R}}-\Delta T$ (with $\Delta T>0$),
we are considering the situation where the heat leaving the hot reservoir can be used to transport charge against the bias voltage, namely when the steady-state electrical and heat currents are
$ J^{\textrm{c}}<0$, and $ J^{\textrm{Q}}_{\textrm{R}}<0$.
This way we can convert waste heat into electrical energy, by charging a battery. 
In such a setup as shown schematically in Fig.~\ref{fig:sketch}, we can define the output power $P$ and efficiency $\eta$ as
\begin{align}
&P=- J^{\textrm{c}} V,\label{eq:outputpower}\\
&\eta=\frac{\mbox{output power}}{\mbox{input heat}}=\frac{P}{- J^{\textrm{Q}}_{\textrm{R}}}.
\label{eq:efficiency}
\end{align}

In the linear response regime, the output power reads
\begin{align}
P=-(eV^2)G(T)[1+S(T)(\Delta T/(eV))], 
\end{align}
exhibiting a maximum at $eV/(k_{\rm B}\Delta T)=-eS(T)/2$. The efficiency at maximum power is $\eta/\eta_{C}=(1/2) ZT_0/(ZT_0+2)$.
\begin{figure}[!htbp]
\centering
\includegraphics[width=1.0\linewidth]{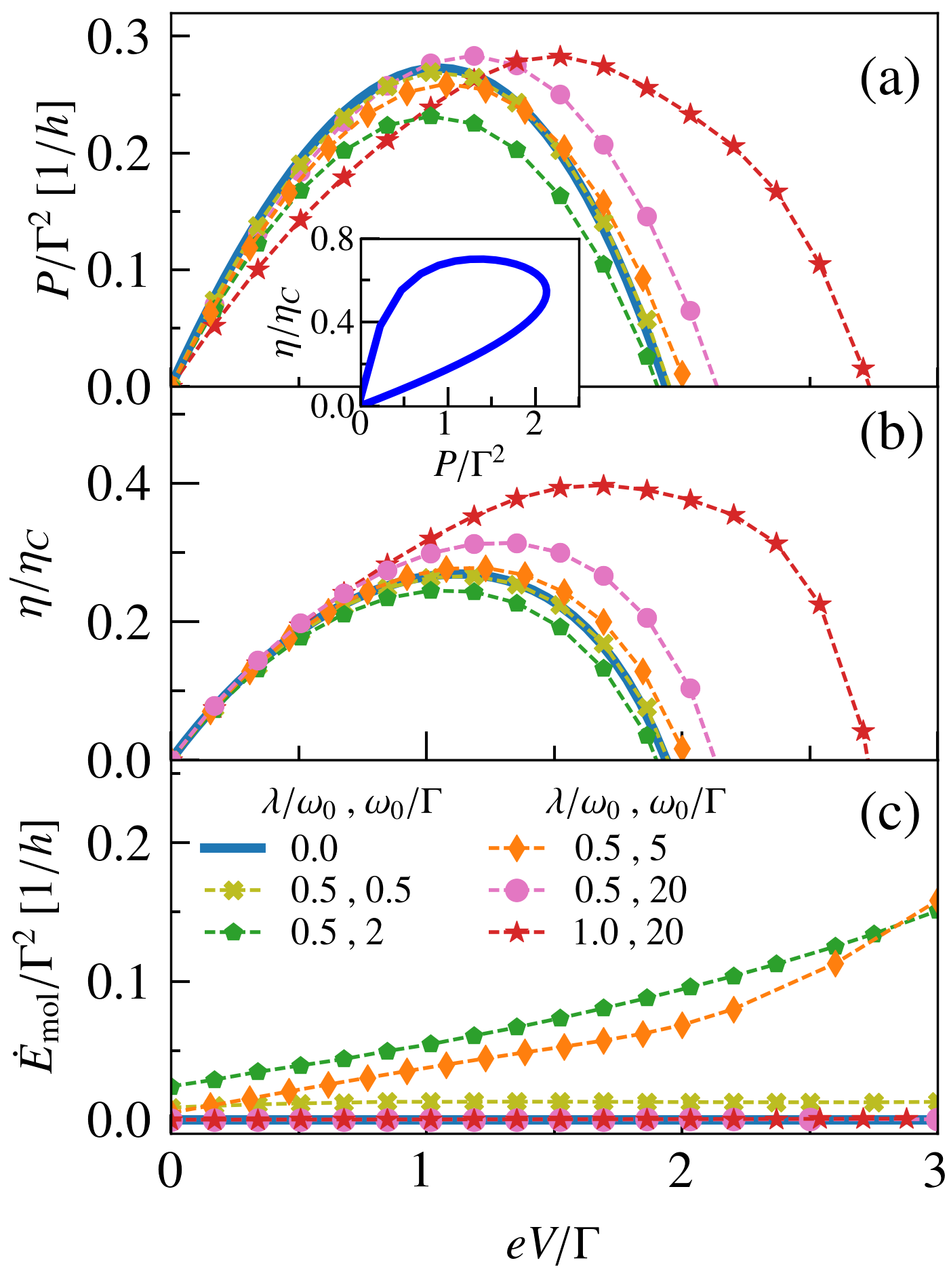}
\caption{(a) Output power, (b) rescaled efficiency, and (c) molecular energy dissipation rate as a function bias voltage for various couplings and phonon frequencies,
at temperatures $k_{\rm B}T_{\rm L}/\Gamma=0.1$, and $k_{\rm B}T_{\rm R}/\Gamma=2.1$, and for gate voltage $\tilde{\varepsilon_0}/\Gamma=2$.
The inset
shows the efficiency as a function of output power for $\lambda=0$, $k_{\rm B}T_{\rm L}/\Gamma=1$, $k_{\rm B}T_{\rm R}/\Gamma=90$, and $\tilde{\varepsilon}_0/\Gamma=40$.}
\label{fig:peta}
\end{figure}
Hence, within linear response, the efficiency at maximum power is bounded above by half of the Carnot efficiency ($\eta_{\rm C}=\Delta T/T_{\rm R}$),
with the upper bound being attained in the ideal situation
of $ZT_0\to\infty$.\cite{Broeck05,Benenti11}
However beyond linear response, the efficiency can go beyond this upper bound 
as has been shown for the IRLM in Ref.~\onlinecite{Kennes13}, and 
is illustrated for a single resonant level model ($\lambda=0$) in the inset to Fig.~\ref{fig:peta} (a).
In this work we are interested to characterize the parameter regimes for which the efficiency can be improved merely
due to the presence of the vibrational degrees of freedom.

Beyond linear response, we discuss the evolution of the output power, and efficiency as a function of bias-voltage as we vary the phonon frequency, and the electron-phonon coupling.
In the adiabatic limit $\omega_0\ll\Gamma_{\rm eff}$ (vibrations are slow compared to charge fluctuations) the behavior is very much similar to the non-interacting case, see Fig.~\ref{fig:peta} (a).
As we increase the phonon frequency further both the output power and efficiency, at any given bias voltage, exhibit a non-monotonic behavior, i.e., they first decrease and then increase.
This observation can be traced back to the frequency dependence of the spectral function, see the Appendix.
As we progressively enter the antiadiabatic regime, the resonant tunneling is suppressed due to the finite gate voltage, and the inelastic scattering processes are becoming more relevant.
However, deep in the antiadiabatic limit ($\omega_0/\Gamma=20$), the satellite peaks are pushed outside the transport window,
and the sharp resonant tunneling not only results in the enhancement of output power and efficiency 
but also extends the parameter regime in which the system can act as a generator.
Figures~\ref{fig:peta} (a) and \ref{fig:peta} (b) indicate that the mentioned effects (in the antiadibatic limit) can be further pronounced, if we increase the electron-phonon coupling.
To analyze the underlying physics, we look at the molecular energy dissipation rate as a function of bias-voltage as shown in Fig.~\ref{fig:peta} (c).
While in the absence of electron-phonon coupling, the molecular dissipation rate vanishes,
at a finite coupling $\lambda\neq 0$, the molecule dissipates energy via the vibrational degrees of freedom.
As shown in Fig.~\ref{fig:peta} (c), increasing the phonon frequency at the fixed coupling $\lambda/\omega_0=0.5$
elevates the molecular dissipation rate.
However, as we approach the antiadiabatic limit $\dot{E}_{\rm mol}$ decreases
(compare the corresponding curve for $\omega_0/\Gamma=2$ with the one for $\omega_0/\Gamma=5$), and eventually  
deep in the antiadiabatic limit, the molecular dissipation rate is quite small,
and it remains comparably small when increasing the electron-phonon coupling further.
As the latter implies that less energy is being dissipated in the phonon bath,
we conclude that the suppression of the molecular dissipation rate in the antiadiabatic limit is crucial for the observed enhancement of the thermoelectric efficiency.
\section{Summary and perspective}
\label{sec:conclusion}
We applied the FRG method to study the (steady-state) thermoelectric transport
through a vibrating molecular quantum dot in the framework of the spinless Anderson Holstein model.
We presented the technical details of employing FRG on the Keldysh contour to a retarded two-particle interaction.
In the linear response regime we provided comparisons of the FRG results to the ones obtained from the NRG approach,
finding good agreement over the whole temperature range for weak to intermediate electron-phonon couplings.
We showed that the first-order truncated FRG (controlled for weak to intermediate electron-phonon coupligs)
can indeed capture the distinct signatures of phonon-assisted tunneling in the current-bias-voltage characteristic beyond linear response.
In particular, we discussed the Franck-Condon blockade, and the phononic steps appearing in the bias voltage dependency of the electrical current and the differential conductance.
Finally, we specified the parameter regime in which vibrational effects can be used to enhance the output power, and efficiency, in the context of
thermoelectric generators.


  
We should emphasize that our study is valid from the low-temperature limit $T\ll\Gamma_{\rm eff}$ to the high-temperature weak-coupling one $T\gg\Gamma_{\rm eff}$.
In the latter regime the real time diagramatics (RTD) has been applied to study the effects of the local electron-electron, and electron-phonon interaction
in the nonlinear electrical and heat conduction.\cite{Leijnse08,Leijnse10,Gergs15}
Recently a quantum-dot heat engine operating based on the thermally driven flow of particles has been experimentally realized.\cite{Linke17} 
However it is challenging to measure the temperatures of the hot/cold electronic leads, and different (dot-lead) tunneling rates.
The RTD have been employed to extract the mentioned parameters
for the 
Coulomb-blockaded
single electron transitor
in Ref.~\onlinecite{Linke17}
and hence to estimate the thermoelectric efficiency $\eta/\eta_{\rm C}\approx70 \%$ at output power of the order of a few $\text{fW}$
(for $\Gamma=5.9~\mu\text{eV}=68~\text{mK}$, $T_{\rm R}=1.54~\text{K}$, and $T_{\rm L}=0.99~\text{K}$).
According to our study 
of a molecular quantum dot 
with vibrational degrees of freedom,
upon using $\Gamma=1~\text{mev}$, the efficiency at maximum power ($\approx 10^4~\text{fW}$) can be improved up to $40\%$
in the antiadiabatic limit $\omega_0/\Gamma=20$ for electron-phonon coupling $\lambda/\omega_0=1$, and $T_{\rm L}=T_{\rm R}/21=1.16~\text{K}$, as shown in Fig.~\ref{fig:peta} (b).

We focused on small to intermediate electron-phonon couplings (which is a limitation of the first-order truncated FRG),
and other than that there was no particular restriction on the parameter space in which the system could be tackled.
In this light our study can be employed as a future reference for the low temperature regime where correlation effects are more pronounced.
The SAHM is a simple model capable of capturing phonon-assisted tunneling in molecular devices.
In a step to make the model more realistic, one can, in future, include the short-range Coulomb interaction at the contact points between the dot and the leads, the electron spin and also a local Coulomb repulsion on the molecule.

\begin{acknowledgments}
 
This work was supported by the Deutscheforschungsgemeinschaft via RTG 1995. We acknowledge
useful discussions with M. R. Wegewijs and supercomputing support
by the John von Neumann institute for Computing (J\"ulich).

\end{acknowledgments}
\appendix*
\begin{figure*}[!htbp]
\centering
\includegraphics[width=1.0\linewidth]{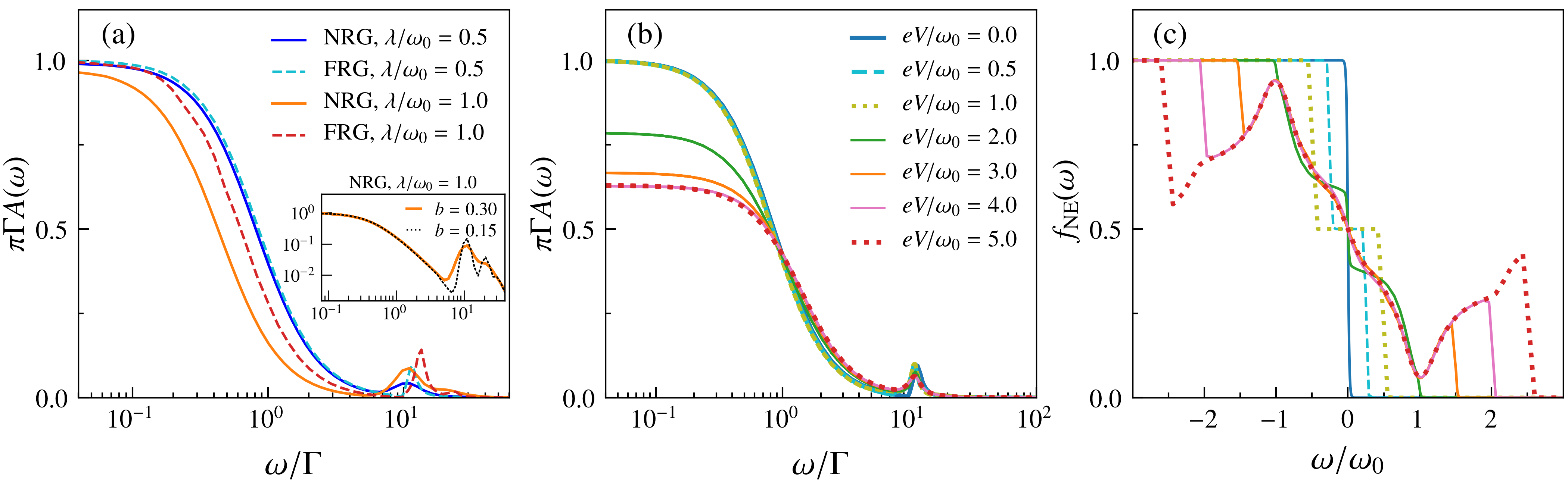}
\caption{(a) The comparison of FRG (dashed lines) and NRG (solid lines) for the frequency dependence of the equilibrium molecular spectral function for two different coupling strengths
in the antiadiabatic limit $\omega_0/\Gamma=10$ at $k_{\rm B}T/\Gamma=0.1$. The inset illustrates $\pi\Gamma A(\omega)$ as a function of $\omega/\Gamma$ within NRG using two different broadening parameters,
for $\lambda/\omega_0=1$.
(b) and (c) The evolution of the frequency dependence of the molecular spectral function,
and the nonequilibrium distribution function Eq.~(\ref{eq:non_dis}) as increasing the bias voltage for $\omega_0/\Gamma=10$, and $k_{\rm B}T/\Gamma=0.1$. Note the logarithmic x-axis in (a) and (b).
In (b) the curve corresponding to $eV/\omega_0=0.5$ is almost indistinguishable from the equilibrium case $eV=0$, and also $eV=4.0$ from $eV/\omega_0=5.0$.}
\label{fig:spec_dis}
\end{figure*}
\section{The molecular spectral and the distribution function}
\label{sec:app_spec}
We discuss the frequency structure of the  molecular spectral and the nonequilibrium distribution function.
First, in the absence of a bias-voltage, we compare the FRG results with the NRG ones.
For weak electron-phonon coupling $\lambda/\omega_0=0.5$, the two methods agree well for low frequencies $\omega<\omega_0$, i.e., for the central quasi-particle peak,
and they only deviate slightly at $\omega\approx\omega_0$ for the satellite side peaks, see Fig.~\ref{fig:spec_dis} (a). 
Within NRG, the spectral function is obtained from the Lehmann representation by broadening the discrete spectra with logarithmic Gaussians,\cite{Bulla08}
and hence the sharpness of the features depends on the broadening parameter used,
as illustrated in the inset of Fig.~\ref{fig:spec_dis} (a).
Therefore on the one hand the features calculated within FRG are sharper which is an artifact of the first order truncation,
and on the other hand the spectral features are more smeared within NRG due to the broadening of the $\delta$ functions (which is a technical necessity).\cite{Bulla08}
However, within NRG the linear thermoelectric transport coefficients at any temperature ($k_{\rm B}T\ll D$) can be directly calculated from the discrete many-body spectrum, \cite{Oliveira09} and hence are highly accurate,
as shown in Fig.~\ref{fig:NTT_GSK_lscan}.

Figure \ref{fig:spec_dis}~(b) shows the evolution of the frequency dependence of the noneqilibrium spectral function upon increasing the bias voltage.
While the modification of the spectral function from its equilibrium value is minor for $eV\ll\omega_0$, for $eV>\omega_0$ the height of the central quasi-particle peak decreases with increasing the bias voltage,
and the features are becoming more broadened.

Figure \ref{fig:spec_dis} (c) shows the variation of the frequency dependence of the nonequilibrium distribution function,
as we increase the bias voltage. It is worth noting that the modifications are significant for bias voltages $eV/2>\omega_0$ in the transport window $\omega\in[-eV/2,eV/2]$.
In particular, at such large bias voltages, we get multiple peaks at $\omega\approx-\omega_0,-2\omega_0,\cdots$,
which suggest that the probability of the states being occupied whenever the energy is sufficient to create one or multiple phonons is enhanced as compared to the effective distribution $f_{\rm eff}(\omega)$,
and analogously the probability of finding the states with frequency $\omega\approx\omega_0,2\omega_0,\cdots$, to be unoccupied are substantially decreased.
\bibliography{ref3}
\end{document}